\documentstyle[ltwol]{article}

\edef\psfigRestoreAt{\catcode`@=\number\catcode`@\relax}
\catcode`\@=11\relax
\newwrite\@unused
\def\typeout#1{{\let\protect\string\immediate\write\@unused{#1}}}
\def\figurepath{./}

\def\@nnil{\@nil}
\def\@empty{}
\def\@psdonoop#1\@@#2#3{}
\def\@psdo#1:=#2\do#3{\edef\@psdotmp{#2}\ifx\@psdotmp\@empty \else
    \expandafter\@psdoloop#2,\@nil,\@nil\@@#1{#3}\fi}
\def\@psdoloop#1,#2,#3\@@#4#5{\def#4{#1}\ifx #4\@nnil \else
       #5\def#4{#2}\ifx #4\@nnil \else#5\@ipsdoloop #3\@@#4{#5}\fi\fi}
\def\@ipsdoloop#1,#2\@@#3#4{\def#3{#1}\ifx #3\@nnil 
       \let\@nextwhile=\@psdonoop \else
      #4\relax\let\@nextwhile=\@ipsdoloop\fi\@nextwhile#2\@@#3{#4}}
\def\@tpsdo#1:=#2\do#3{\xdef\@psdotmp{#2}\ifx\@psdotmp\@empty \else
    \@tpsdoloop#2\@nil\@nil\@@#1{#3}\fi}
\def\@tpsdoloop#1#2\@@#3#4{\def#3{#1}\ifx #3\@nnil 
       \let\@nextwhile=\@psdonoop \else
      #4\relax\let\@nextwhile=\@tpsdoloop\fi\@nextwhile#2\@@#3{#4}}
\newread\ps@stream
\newif\ifnot@eof       
\newif\if@noisy        
\newif\if@atend        
\newif\if@psfile       
{\catcode`\%=12\global\gdef\epsf@start{
\def\epsf@PS{PS}
\def\epsf@getbb#1{%
\openin\ps@stream=#1
\ifeof\ps@stream\typeout{Error, File #1 not found}\else
   {\not@eoftrue \chardef\other=12
    \def\do##1{\catcode`##1=\other}\dospecials \catcode`\ =10
    \loop
       \if@psfile
	  \read\ps@stream to \epsf@fileline
       \else{
	  \obeyspaces
          \read\ps@stream to \epsf@tmp\global\let\epsf@fileline\epsf@tmp}
       \fi
       \ifeof\ps@stream\not@eoffalse\else
       \if@psfile\else
       \expandafter\epsf@test\epsf@fileline:. \\%
       \fi
          \expandafter\epsf@aux\epsf@fileline:. \\%
       \fi
   \ifnot@eof\repeat
   }\closein\ps@stream\fi}%
\long\def\epsf@test#1#2#3:#4\\{\def\epsf@testit{#1#2}
			\ifx\epsf@testit\epsf@start\else
\typeout{Warning! File does not start with `\epsf@start'.  It may not be a PostScript file.}
			\fi
			\@psfiletrue} 
{\catcode`\%=12\global\let\epsf@percent=
\long\def\epsf@aux#1#2:#3\\{\ifx#1\epsf@percent
   \def\epsf@testit{#2}\ifx\epsf@testit\epsf@bblit
	\@atendfalse
        \epsf@atend #3 . \\%
	\if@atend	
	   \if@verbose{
		\typeout{psfig: found `(atend)'; continuing search}
	   }\fi
        \else
        \epsf@grab #3 . . . \\%
        \not@eoffalse
        \global\no@bbfalse
        \fi
   \fi\fi}%
\def\epsf@grab #1 #2 #3 #4 #5\\{%
   \global\def\epsf@llx{#1}\ifx\epsf@llx\empty
      \epsf@grab #2 #3 #4 #5 .\\\else
   \global\def\epsf@lly{#2}%
   \global\def\epsf@urx{#3}\global\def\epsf@ury{#4}\fi}%
\def\epsf@atendlit{(atend)} 
\def\epsf@atend #1 #2 #3\\{%
   \def\epsf@tmp{#1}\ifx\epsf@tmp\empty
      \epsf@atend #2 #3 .\\\else
   \ifx\epsf@tmp\epsf@atendlit\@atendtrue\fi\fi}

\def\psdraft{
	\def\@psdraft{0}
}
\def\psfull{
	\def\@psdraft{100}
}

\psfull

\newif\if@draftbox
\def\psnodraftbox{
	\@draftboxfalse
}
\@draftboxtrue

\newif\if@prologfile
\newif\if@postlogfile
\def\pssilent{
	\@noisyfalse
}
\def\psnoisy{
	\@noisytrue
}
\psnoisy
\newif\if@bbllx
\newif\if@bblly
\newif\if@bburx
\newif\if@bbury
\newif\if@height
\newif\if@width
\newif\if@rheight
\newif\if@rwidth
\newif\if@clip
\newif\if@verbose
\def\@p@@sclip#1{\@cliptrue}

\def\@p@@sfile#1{\def\@p@sfile{null}%
	        \openin1=#1
		\ifeof1\closein1%
		       \openin1=\figurepath#1
			\ifeof1\typeout{Error, File #1 not found}
			\else\closein1
			    \edef\@p@sfile{\figurepath#1}%
                        \fi%
		 \else\closein1%
		       \def\@p@sfile{#1}%
		 \fi}
\def\@p@@sfigure#1{\def\@p@sfile{null}%
	        \openin1=#1
		\ifeof1\closein1%
		       \openin1=\figurepath#1
			\ifeof1\typeout{Error, File #1 not found}
			\else\closein1
			    \def\@p@sfile{\figurepath#1}%
                        \fi%
		 \else\closein1%
		       \def\@p@sfile{#1}%
		 \fi}

\def\@p@@sbbllx#1{
		\@bbllxtrue
		\dimen100=#1
		\edef\@p@sbbllx{\number\dimen100}
}
\def\@p@@sbblly#1{
		\@bbllytrue
		\dimen100=#1
		\edef\@p@sbblly{\number\dimen100}
}
\def\@p@@sbburx#1{
		\@bburxtrue
		\dimen100=#1
		\edef\@p@sbburx{\number\dimen100}
}
\def\@p@@sbbury#1{
		\@bburytrue
		\dimen100=#1
		\edef\@p@sbbury{\number\dimen100}
}
\def\@p@@sheight#1{
		\@heighttrue
		\dimen100=#1
   		\edef\@p@sheight{\number\dimen100}
}
\def\@p@@swidth#1{
		\@widthtrue
		\dimen100=#1
		\edef\@p@swidth{\number\dimen100}
}
\def\@p@@srheight#1{
		\@rheighttrue
		\dimen100=#1
		\edef\@p@srheight{\number\dimen100}
}
\def\@p@@srwidth#1{
		\@rwidthtrue
		\dimen100=#1
		\edef\@p@srwidth{\number\dimen100}
}
\def\@p@@ssilent#1{ 
		\@verbosefalse
}
\def\@p@@sprolog#1{\@prologfiletrue\def\@prologfileval{#1}}
\def\@p@@spostlog#1{\@postlogfiletrue\def\@postlogfileval{#1}}
\def\@cs@name#1{\csname #1\endcsname}
\def\@setparms#1=#2,{\@cs@name{@p@@s#1}{#2}}
\def\ps@init@parms{
		\@bbllxfalse \@bbllyfalse
		\@bburxfalse \@bburyfalse
		\@heightfalse \@widthfalse
		\@rheightfalse \@rwidthfalse
		\def\@p@sbbllx{}\def\@p@sbblly{}
		\def\@p@sbburx{}\def\@p@sbbury{}
		\def\@p@sheight{}\def\@p@swidth{}
		\def\@p@srheight{}\def\@p@srwidth{}
		\def\@p@sfile{}
		\def\@p@scost{10}
		\def\@sc{}
		\@prologfilefalse
		\@postlogfilefalse
		\@clipfalse
		\if@noisy
			\@verbosetrue
		\else
			\@verbosefalse
		\fi
}
\def\parse@ps@parms#1{
	 	\@psdo\@psfiga:=#1\do
		   {\expandafter\@setparms\@psfiga,}}
\newif\ifno@bb
\def\bb@missing{
	\if@verbose{
		\typeout{psfig: searching \@p@sfile \space  for bounding box}
	}\fi
	\no@bbtrue
	\epsf@getbb{\@p@sfile}
        \ifno@bb \else \bb@cull\epsf@llx\epsf@lly\epsf@urx\epsf@ury\fi
}	
\def\bb@cull#1#2#3#4{
	\dimen100=#1 bp\edef\@p@sbbllx{\number\dimen100}
	\dimen100=#2 bp\edef\@p@sbblly{\number\dimen100}
	\dimen100=#3 bp\edef\@p@sbburx{\number\dimen100}
	\dimen100=#4 bp\edef\@p@sbbury{\number\dimen100}
	\no@bbfalse
}
\def\compute@bb{
		\no@bbfalse
		\if@bbllx \else \no@bbtrue \fi
		\if@bblly \else \no@bbtrue \fi
		\if@bburx \else \no@bbtrue \fi
		\if@bbury \else \no@bbtrue \fi
		\ifno@bb \bb@missing \fi
		\ifno@bb \typeout{FATAL ERROR: no bb supplied or found}
			\no-bb-error
		\fi
		\count203=\@p@sbburx
		\count204=\@p@sbbury
		\advance\count203 by -\@p@sbbllx
		\advance\count204 by -\@p@sbblly
		\edef\@bbw{\number\count203}
		\edef\@bbh{\number\count204}
}
\def\in@hundreds#1#2#3{\count240=#2 \count241=#3
		     \count100=\count240	
		     \divide\count100 by \count241
		     \count101=\count100
		     \multiply\count101 by \count241
		     \advance\count240 by -\count101
		     \multiply\count240 by 10
		     \count101=\count240	
		     \divide\count101 by \count241
		     \count102=\count101
		     \multiply\count102 by \count241
		     \advance\count240 by -\count102
		     \multiply\count240 by 10
		     \count102=\count240	
		     \divide\count102 by \count241
		     \count200=#1\count205=0
		     \count201=\count200
			\multiply\count201 by \count100
		 	\advance\count205 by \count201
		     \count201=\count200
			\divide\count201 by 10
			\multiply\count201 by \count101
			\advance\count205 by \count201
		     \count201=\count200
			\divide\count201 by 100
			\multiply\count201 by \count102
			\advance\count205 by \count201
		     \edef\@result{\number\count205}
}
\def\compute@wfromh{
		\in@hundreds{\@p@sheight}{\@bbw}{\@bbh}
		\edef\@p@swidth{\@result}
}
\def\compute@hfromw{
		\in@hundreds{\@p@swidth}{\@bbh}{\@bbw}
		\edef\@p@sheight{\@result}
}
\def\compute@handw{
		\if@height 
			\if@width
			\else
				\compute@wfromh
			\fi
		\else 
			\if@width
				\compute@hfromw
			\else
				\edef\@p@sheight{\@bbh}
				\edef\@p@swidth{\@bbw}
			\fi
		\fi
}
\def\compute@resv{
		\if@rheight \else \edef\@p@srheight{\@p@sheight} \fi
		\if@rwidth \else \edef\@p@srwidth{\@p@swidth} \fi
}
\def\compute@sizes{
	\compute@bb
	\compute@handw
	\compute@resv
}
\def\psfig#1{\vbox {
	\ps@init@parms
	\parse@ps@parms{#1}
	\compute@sizes
	\ifnum\@p@scost<\@psdraft{
		\if@verbose{
			\typeout{psfig: including \@p@sfile \space }
		}\fi
		\special{ps::[begin] 	\@p@swidth \space \@p@sheight \space
				\@p@sbbllx \space \@p@sbblly \space
				\@p@sbburx \space \@p@sbbury \space
				startTexFig \space }
		\if@clip{
			\if@verbose{
				\typeout{(clip)}
			}\fi
			\special{ps:: doclip \space }
		}\fi
		\if@prologfile
		    \special{ps: plotfile \@prologfileval \space } \fi
		\special{ps: plotfile \@p@sfile \space }
		\if@postlogfile
		    \special{ps: plotfile \@postlogfileval \space } \fi
		\special{ps::[end] endTexFig \space }
		\vbox to \@p@srheight true sp{
			\hbox to \@p@srwidth true sp{
				\hss
			}
		\vss
		}
	}\else{
		\if@draftbox{		
			\hbox{\fbox{\vbox to \@p@srheight true sp{
			\vss
			\hbox to \@p@srwidth true sp{ \hss \@p@sfile \hss }
			\vss
			}}}
		}\else{
			\vbox to \@p@srheight true sp{
			\vss
			\hbox to \@p@srwidth true sp{\hss}
			\vss
			}
		}\fi

	}\fi
}}
\def\psglobal{\typeout{psfig: PSGLOBAL is OBSOLETE; use psprint -m instead}}
\psfigRestoreAt

\arraycolsep1.5pt 

\def\Journal#1#2#3#4{{#1} {\bf #2}, #3 (#4)}

\def\NCA{\em Nuovo Cimento}
\def\NIM{\em Nucl. Instrum. Methods}
\def\NIMA{{\em Nucl. Instrum. Methods} A}
\def\NPB{{\em Nucl. Phys.} B}
\def\PLB{{\em Phys. Lett.}  B}
\def\PRL{\em Phys. Rev. Lett.}
\def\PRD{{\em Phys. Rev.} D}
\def\ZPC{{\em Z. Phys.} C}

\def\st{\scriptstyle}
\def\sst{\scriptscriptstyle}
\def\mco{\multicolumn}
\def\epp{\epsilon^{\prime}}
\def\vep{\varepsilon}
\def\ra{\rightarrow}
\def\ppg{\pi^+\pi^-\gamma}
\def\vp{{\bf p}}
\def\ko{K^0}
\def\kb{\bar{K^0}}
\def\al{\alpha}
\def\ab{\bar{\alpha}}
\def\be{\begin{equation}}
\def\ee{\end{equation}}
\def\bea{\begin{eqnarray}}
\def\eea{\end{eqnarray}}
\def\CPbar{\hbox{{\rm CP}\hskip-1.80em{/}}}

\def\atwo       {{\mbox{$a_{2}                                        \ $}}}
\def\a2         {{\mbox{$a_{2}                                        \ $}}}
\def\piplus     {{\mbox{$\pi^{+}                                      \ $}}}
\def\piminus    {{\mbox{$\pi^{-}                                      \ $}}}
\def\pizero     {{\mbox{$\pi^{0}                                      \ $}}}
\def\ptsq       {{\mbox{$p_T^{2}                                      \ $}}}
\def\gevcc      {{\mbox{$(GeV/c)^{2}                                  \ $}}} 
\def\a2pigamma  {{\mbox{$a_2^-\rightarrow \pi^-\gamma                \ $}}}
\def\gapigamma  {{\mbox{$\Gamma(a_2^-\rightarrow \pi^-\gamma)        \ $}}}
\def\reacta     {{\mbox{$\pi^-+A\rightarrow A+(\pi^+\pi^-\pi^-)      \ $}}}
\def\reactcb    {{\mbox{$\pi^-+C\rightarrow C+(\pi^+\pi^-\pi^-)      \ $}}}
\def\reactcu    {{\mbox{$\pi^-+Cu\rightarrow Cu+(\pi^+\pi^-\pi^-)      \ $}}}
\def\reactpb    {{\mbox{$\pi^-+Pb\rightarrow Pb+(\pi^+\pi^-\pi^-)      \ $}}}
\def\threepi    {{\mbox{$(\pi^+\pi^-\pi^-)                             \ $}}}

\bibliographystyle{unsrt}    

\begin{document}

\title{Radiative Width of the $a_2$ Meson$^\dag $}



%
%
%
%
%
%
\author{\large The \textsc{Selex} Collaboration\\}
\author{\small\noindent
  V.P.~Kubarovsky$^{6}$,
  N.~Akchurin$^{17}$,
  V.~A.~Andreev$^{11}$,
  A.G.~Atamantchouk$^{11}$,
  M.~Aykac$^{17}$,
  M.Y.~Balatz$^{8}$,
  N.F.~Bondar$^{11}$,
  A.~Bravar$^{22}$,
  M.~Chensheng$^{7}$,
  P.S.~Cooper$^{5}$
% ...
\\
}

\address{\noindent\footnotesize
% ...
$^{5}$ Fermilab, Batavia, IL 60510, U.S.A.\\
$^{6}$ Institute for High Energy Physics, Protvino, Russia\\
$^{7}$ Institute of High Energy Physics, Beijing, PR China\\
$^{8}$ Institute of Theoretical and Experimental Physics, Moscow, 
Russia\\
% ...
$^{11}$ Petersburg Nuclear Physics Institute, St. Petersburg, Russia\\
% ...
$^{17}$ University of Iowa,  Iowa City, Iowa  52242, U.S.A.\\
% ...
$^{22}$ University of Trieste and INFN, Trieste, Italy\\
}



\twocolumn[\maketitle\abstracts{
We present data on coherent production of
the \threepi system by 600 GeV/c pion beam in the reaction
\reacta for the C, Cu and Pb targets.
The Primakoff formalism was
used for extracting the radiative width of the \atwo meson. We obtain
a preliminary value 
$\Gamma ( \a2pigamma )=225\pm25({\rm stat})\pm45({\rm syst})$ keV. 
}]


\section{Introduction}

Radiative decay widths of mesons and baryons are powerful tools for 
understanding  the structure of  elementary particles and for constructing
a dynamical theory of hadronic systems. Straightforward predictions for
radiative widths make possible the direct comparison of experimental data
and theory.

The small value of the branching ratios of radiative decays makes it difficult
to measure them directly, 
if there is a large background from the strong
decay channels with $\pi^0\to\gamma\gamma$
in the final state with one lost photon.
For this reason most experimental data
for these decays have been obtained using
production reactions in the Coulomb field of the nuclei.
Studying the inverse reaction 
(the so called Primakoff \cite{primak} formalism)
\mbox{$\gamma+\pi^-\rightarrow M^-$}
provides a relatively clean method for the determination of the radiative
widths.

The Primakoff reaction works better when the energy of the beam
particles is increased. However, very good spatial resolution is needed to
extract the signal at  small $t$ where the electromagnetic processes
dominate over the strong interaction.

SELEX (E781) had a beam energy of $600$ GeV and had 
a high resolution
vertex detector which made it possible to explore the features of
the Primakoff reaction.

The differential cross section for the Primakoff reaction 
\begin{equation}
\label{frm6}
\pi^-+Z\to Z+a^-_2,~~a_2^-\to \pi^-\pi^-\pi^+
\end{equation}
can be written as a function of the mass of the final state ($M$) and 
the square of the four-momentum transfer to the nucleus ($t$) as follows

\begin{eqnarray}
\label{frm8}
&\frac{d\sigma}{dt~dM} =
8\pi\alpha Z^2(2J_{a_2}+1)\Gamma_{\gamma} 
\left( \frac{M}{M^2-m_{\pi}^2} \right)^3
g_{\gamma}(M)g_{a_2}(M) 
\nonumber \\
&\frac{2}{\pi}
\frac{m^2_{a_2}\Gamma_{a_2}}{(M^2-m^2_{a_2})^2+
m^2_{a_2}g^2_{a_2}(M)\Gamma^2_{a_2}}
\frac{t-t_{min}}{t^2} |F(t)|^2 
\end{eqnarray}
Here $\alpha$ is the fine-structure constant, 
$Z$ is the charge of the nucleus,
$F(t)$ is the electromagnetic form factor of the nucleus,
$J_{a_2}$ is the spin of the $a_2$ meson and  
$t_{min}$ is the minimum four-momentum transfer:
\begin{equation}
t_{min}\approx(M^2-m_\pi^2)^2/4P_{beam}^2, 
\end{equation}
where $P_{beam}$ is the beam momentum. In our case $P_{beam}=600$ GeV and
for the $a_2$ meson with the mass $1.32$~GeV we have
\begin{equation}
t_{min}\approx 2\cdot10^{-6} GeV^2
\end{equation}
The $t$-distribution for the Primakoff reaction  
is characterized by a sharp peak with the
maximum near $t=2\cdot t_{min}$. The influence of the nuclear form factor is
small in the region of $t<0.001$ GeV$^2$ where we expect the
Primakoff signal. It is less than 5\% for the Pb target and 3\% 
for the Cu target.

The mass-dependent total width and the partial widths  
$\Gamma(a_2\to all)$, 
$\Gamma(a_2\to\pi\gamma)$ and
$\Gamma(a_2\to\rho\pi)$
can be
rewritten in terms of resonance widths as follows: 
$\Gamma(a_2\to\pi\gamma)=\Gamma_{\gamma}g_{\gamma}(M)$ and
$\Gamma(a_2\to\rho\pi)=\Gamma_{\rho}g_{\rho}(M)$, where
$g_{\gamma}(m_{a_2})=g_{\rho}(m_{a_2})=1$. We used 
$g(a_2\to all)=g(a_2\to \rho\pi)$.

For the energy dependence of the widths we used
\begin{eqnarray}
\label{frm9}
g_\gamma(M)=\left( \frac{k}{k_0} \right)^3 \frac{2k_0^2}{k^2+k^2_0}\\
g_{a_2}(M)=\left( \frac{q}{q_0} \right)^5 \frac{2q_0^2}{q^2+q^2_0} 
\end{eqnarray}
where $k$ and $q$ are the momenta in the $\pi\gamma$ and $\rho\pi$
frames, respectively. 

Numerically integrating over $M$ and $t$ from $t_{min}$ up to $0.001$~GeV$^2$ 
we get the cross section for the reaction (\ref{frm6}) in this $t$-interval
\begin{equation}
\label{frm11}
\sigma_{\rm Primakoff}=
\int \limits_{t_{min}}^{0.001}\frac{d\sigma}{dt~dM}dt~dM=
\Gamma(a_2\to\pi\gamma)\cdot C,
\end{equation}
where $C=27.2,~626,~4870~(mb/GeV)$ for the C, Cu and Pb targets
respectively.

\section{Cuts and Statistics of the Experiment}

The following cuts were applied in order to extract the reaction
\begin{equation}
\label{frm12}
\pi^- + A \to A + (\pi^-\pi^-\pi^+)
\end{equation}
from the exclusive trigger stream of SELEX data:

\begin{enumerate}
\item{}
The beam transition radiation detector shows that the beam 
particle is a pion.
\item{}
For the quasielastic peak, the difference between the beam particle 
momentum and the
sum of momenta of three secondaries must be less than 
$17.5$~GeV/c.
\item{}
The $z$-position of the interaction must be in the vicinity of 
a C, Cu or Pb target.
\item{}
The energy deposition in the Photon-1 detector must be less than 2 GeV.
\item{}
The average distance between the point of interaction and all tracks (beam
and secondaries) must be less than 20, 25 and 75 $\mu$m for 
the C, Cu and Pb
targets.

\end{enumerate}

The event statistics are presented in Table~\ref{ta:stat}.

\begin{table}
\begin{center}
\caption{Statistics of the experiment.}
\label{ta:stat}
\begin{tabular}{|c|r|}
\hline
Target & Number of Events \\
\hline
\hline
Carbon &  2~760~523\\
\hline
Copper & 1~997~972\\
\hline
Lead   &  549~092\\
\hline
\end{tabular}
\end{center}
\end{table}

\section{Primakoff Production of the a2 Meson}

\begin{figure*}
\centerline{\psfig{figure=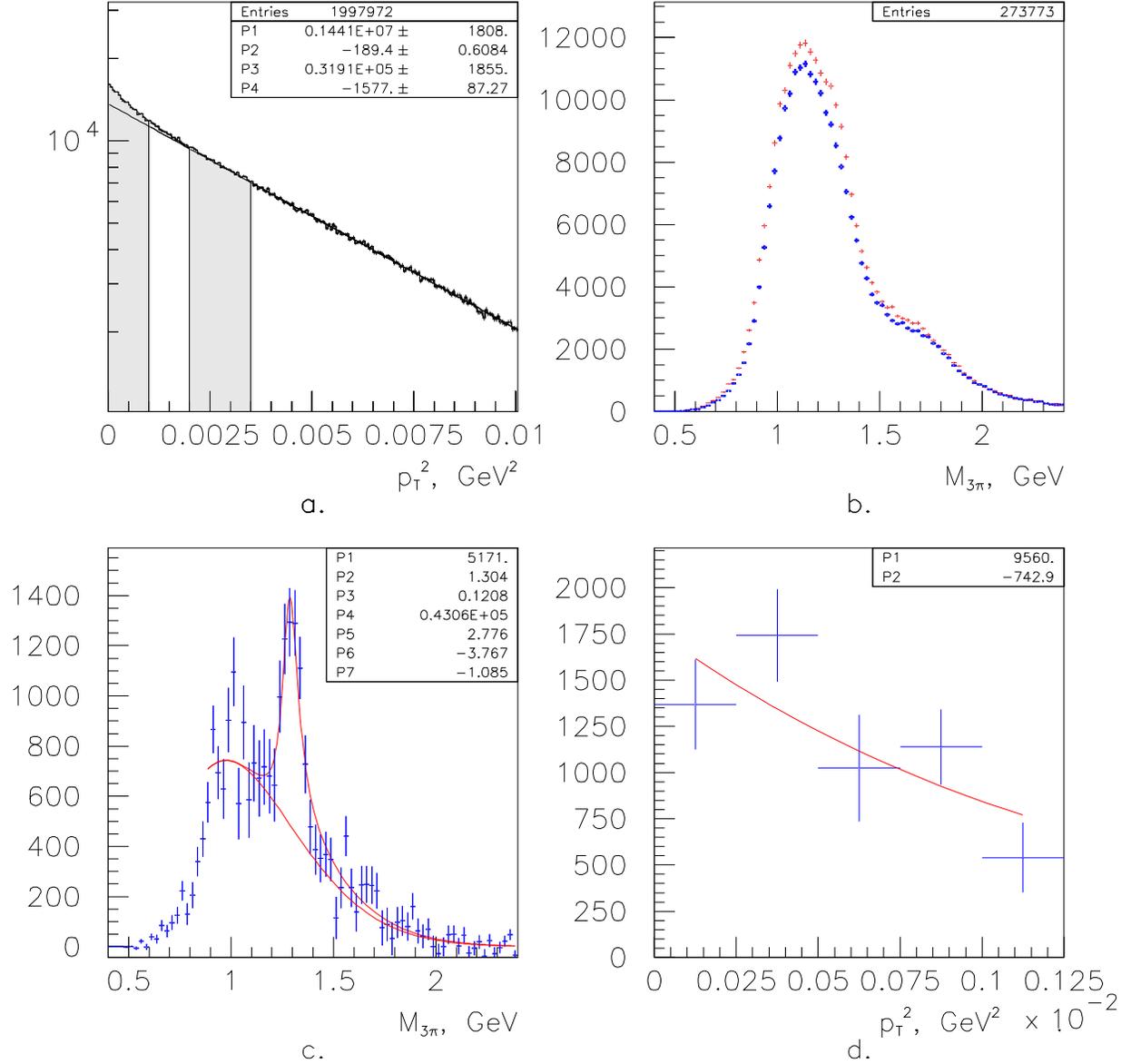,height=18cm,width=18cm} }
\caption{Primakoff production of the 
\atwo meson on Copper target in the reaction:
\mbox{$\pi^- + Cu \rightarrow Cu + ( \pi^-\pi^+\pi^-)$}.
a. \ptsq distribution of $(3\pi)$-system. The Primakoff signal is clearly
seen in the region of small \ptsq.
b. $M_{3\pi}$ distribution. The first curve(light points) is for events
with $\ptsq<0.001$ GeV$^2$, the second one is for events in the band 
\mbox{$0.002<\ptsq<0.0035$ GeV$^2$}. 
c. Result of subtraction of two curves (see b). 
d. \ptsq distribution of the \atwo meson.}
\label{fig:cu}
\end{figure*}

The \ptsq 
distribution of the $(3\pi)$-system for the Cu target is shown in
Fig.~\ref{fig:cu}a as an example. The distribution was fitted by the sum of
two exponentials. The slope for the coherent production was found to be
close to the
previously published data \cite{ferbell3pi}.
The \ptsq distributions for all targets have an enhancement in the region of
small \ptsq. 
The two exponentials fit yields a 
slope for the second term greater
than 1000~GeV$^{-2}$ which corresponds to the Monte Carlo estimate for
the Primakoff reaction. 
 
Two $\ptsq$ regions were defined (the shaded regions in 
Fig.~\ref{fig:cu}a). 
In the first one
\mbox{($\ptsq<0.001$~GeV$^2$)} the Primakoff mechanism dominates.
We used the second one
\mbox{$0.0015<\ptsq<0.0035$~GeV$^2$}
for the estimation of the background from the coherent production 
of the $(3\pi)$-system. 

The $\pi\pi\pi$ mass spectra for these two regions are shown 
in  
Fig.~\ref{fig:cu}b 
for the Cu target.
The result of the subtraction of these histograms is
shown in 
Fig.~\ref{fig:cu}c. 
A clear $a_2$ meson signal is seen. The distributions were 
fitted using Eq.~(\ref{frm8}) integrated
over $t$.
The mass and width of the resonance were determined using the
combined data from all targets: $M_{a_2}=1304\pm 5$~MeV, 
$\Gamma_{a_2}=121\pm 20$~MeV.
The mass and the width of 
the peaks are in 
good agreement with the mass and width of the \atwo meson. 
The number of $a_2$ events was found to be
$1587\pm 480$, $5170\pm 590$ and $2945\pm 400$ for the C, Cu an Pb 
targets, respectively.

To make sure that the $a_2$ meson is produced via the Primakoff mechanism 
we divided the data into five \ptsq-bins each
0.00025~GeV$^2$ and repeated
the subtraction procedure. We got 5 histograms for each target similar to 
Fig.~\ref{fig:cu}c. 
After the fit of the mass spectra and the 
determination the
number of \atwo mesons in each bin the \ptsq distribution was extracted
and shown
in 
Fig.~\ref{fig:cu}d. 
The slope was determined for every target. It is
consistent with the Primakoff distribution smeared by the resolution of the
experimental setup.

\section{Monte Carlo}

A simple Monte Carlo program was written to simulate 
the \ptsq -distribution for the Primakoff production of
the \atwo meson.
The $p_T$ resolution had these  different
values for the different targets:
$\sigma(p_{Tx})=\sigma (p_{Ty})=16,~17$ and 20~MeV for 
the C, Cu and Pb 
targets, respectively.

The subtraction procedure efficiency was
evaluated using the Monte Carlo simulation:
\begin{equation}
\label{frm15}
\varepsilon(subtr.)=
\frac
{  \int \limits_0^{0.001}        \frac{d\sigma'}{dt}dt-k
   \int \limits_{0.0015}^{0.0035}\frac{d\sigma'}{dt}dt  }
{  \int \limits_0^{0.001}\frac{d\sigma}{dt}dt}
\end{equation}
where $\sigma$ is the Primakoff cross section, $\sigma'$ is the
Primakoff cross section 
smeared by
the experimental resolution
and $k$ is the normalization factor.
These efficiencies were found to be
$\varepsilon=0.697,~0.630$ and $0.488$ for the C, Cu and Pb targets.

\section{Absolute Normalization}

The Primakoff approach gives the possibility to determine the radiative width 
only in the case when the absolute cross section of 
the reaction (\ref{frm6}) is
measured. This is the crucial point for the experiment. At 
the present stage of
analysis we use 
the cross section for the reaction (\ref{frm12}) from 
E272~\cite{ferbell3pi}. We determined the product of
($L\cdot\varepsilon$) and didn't need to evaluate the reconstruction efficiency
$\varepsilon$ alone. 
Here $L$ is the luminosity.
The experiment E272 
was done in a beam with energy 200~GeV.
We assumed that the coherent cross section of the reaction (\ref{frm12}) 
is independent
of the beam energy. The simplest estimation based on Regge theory 
confirms this assumption. 

We determined the absolute cross section in the \ptsq region
close to zero to avoid the acceptance corrections:
\begin{equation}
\label{frm14}
(L\cdot \varepsilon)=
\frac
{\int \limits_{t_{min}}^{t_{max}}\frac{dN}{dt}dt}
{\int \limits_{t_{min}}^{t_{max}}\frac{d\sigma}{dt}dt}
\end{equation}
where 
$\frac{dN}{dt}$ is the differential t-distribution from SELEX data,
$\frac{d\sigma}{dt}$ is the differential cross section from E272 data,
$t_{min}=0,~0.001,~0.0015$~GeV$^2$ and
$t_{max}=0.020,~0.020,~0.006$~GeV$^2$  for three targets.
$(L\cdot\varepsilon)=881558,~149302,~17831~~$events/mb for 
the C, Cu and Pb targets, respectively.

\section{Radiative Width of the \atwo Meson}

\begin{figure}[ht]
\centerline{\psfig{figure=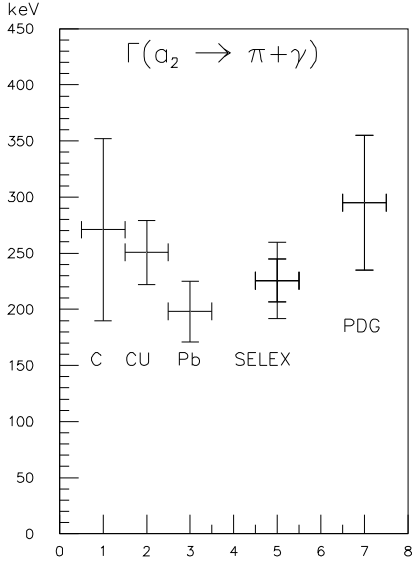,width=8cm} }
\caption{The radiative width 
$\Gamma ( \a2pigamma ) $ for the C, Cu and Pb targets. The
average of SELEX result 
(with statistical and systematic errors) 
and the PDG value are shown as well.}
\label{fig:width}
\end{figure}

The radiative width of the  \atwo meson was determined using the expression
\begin{equation}
\label{frm16}
\gapigamma=\frac
{N_{\atwo}}
{C\cdot(L\varepsilon)\cdot CG\cdot BR
\cdot\varepsilon(subtr.)}
\end{equation}
where 
$CG(\atwo\to\rho\pi)$=0.5 is the Clebsch-Gordon coefficient,
$BR(\atwo\to\rho\pi)$=0.701 is the branching ratio,
$C$ is determined by numerical integration (see Eq.~(\ref{frm11})),
$(L\varepsilon)$ is the product of the luminosity and the reconstruction
efficiency (see Eq.~(\ref{frm14})). The radiative width was determined for
all three targets: C, Cu and Pb. 
The results are shown in 
Fig.~\ref{fig:width}. 
As can be seen
from this figure the results are consistent with each other,  confirming
the Coulomb production of the \atwo meson. The average 
value over all targets is
\begin{equation}
\label{frm17}
\gapigamma=225\pm 20({\rm stat})\pm 45({\rm syst})~{\rm keV}
\end{equation}
The major sources of  
uncertainty in this result are the normalization procedure
\cite{ferbell3pi}
and the errors in the determination of the number of \atwo events.
The PDG value for the $\gapigamma=295\pm60$ keV
is shown in 
Fig.~\ref{fig:width} as well.

\section{Conclusion}

Based on a weighted average over the C, Cu and Pb targets 
of the values for \gapigamma we present the
preliminary value of
$225\pm 20({\rm stat})\pm 45({\rm syst})$ keV 
for the radiative width of the \atwo 
meson.
Our result has the best world statistical error \cite{cihangir,pdg}
for the radiative width of
the \atwo meson.
The systematic error can be reduced in the 
future by accurate measurement of the
\threepi coherent cross section  using SELEX data.

\footnotetext[0]{\noindent $^\dag$Talk presented at the 1998
International Conference on High Energy Physics,
Vancouver, B.C., Canada, July, 1998.}
\addtocounter{footnote}{-12}
\addtocounter{footnote}{1}
\footnotetext{Present address: Instituto
    de Fisica da Universidade Estadual de Campinas, UNICAMP, SP, 
    Brazil.}
\addtocounter{footnote}{1}
\footnotetext{Now at Imperial College, London SW7 2BZ, U.K.}
\addtocounter{footnote}{1}
\footnotetext{deceased}
\addtocounter{footnote}{1}
\footnotetext{Present address: Dept. of Physics,
      Wayne State University, Detroit, MI 48201, U.S.A.}
\addtocounter{footnote}{1}
\footnotetext{Present address: Universit\"at Freiburg, 79104 Freiburg, 
Germany}
\addtocounter{footnote}{1}
\footnotetext{Present address:
      Physik-Department, Technische Universit\"at M\"unchen,
      85748 Garching, Germany}
\addtocounter{footnote}{1}
\footnotetext{Current Address:
      Instituto de Fisica Teorica da Universidade Estadual Paulista,
      S\~ao Paulo, Brazil}
\addtocounter{footnote}{1}
\footnotetext{Present address: Lucent Technologies,Naperville, IL}
\addtocounter{footnote}{1}
\footnotetext{Now at Max-Planck-Institut f\"ur Physik, M\"unchen, 
Germany}
\addtocounter{footnote}{1}
\footnotetext{Present address: Motorola Inc., Schaumburg, IL}
\addtocounter{footnote}{1}
\footnotetext{Generous support of Carnegie-Mellon University
              is gratefully acknowledged.}
\addtocounter{footnote}{1}
\footnotetext{Present address: Deutsche Bank AG, 65760 Eschborn, 
Germany}


\section*{References}
\begin{thebibliography}{99}

\bibitem{primak}  H.Primakoff, {\em Phys. Rev.} {\bf 81}, 899 (1951).
\bibitem{ferbell3pi}  M.Zielinski $et~al.$, 
{\em Z. Phys.} C {\bf 16}, 197 (1983).
\bibitem{cihangir}  S.Cihangir $et~al.$, {\em Phys. Lett.} B {\bf 117}, 
119 (1982).
\bibitem{pdg}  Particle Data Group, {\em Phys. Rev.} D {\bf 54}, (1996)

\end{thebibliography}
\end{document}